\documentclass[pra,final,twocolumn,amsmath,amssymb,amsfonts]{revtex4-1}
\usepackage{graphicx,graphics}
\usepackage{color}
\usepackage{gensymb}
\usepackage{epsfig}
\usepackage{amsmath}
\usepackage{amssymb}
\usepackage{bm}

\usepackage{textcomp}

\begin{document}

\title{      Spin-orbital order in LaMnO$_3$: $d-p$ model study}

\author{     Krzysztof Ro\'sciszewski}
\affiliation{\mbox{Marian Smoluchowski Institute of Physics, Jagiellonian University,
             Prof. S. \L{}ojasiewicza 11, PL-30348 Krak\'ow, Poland}}

\author{     Andrzej M. Ole\'s  }
\affiliation{Max Planck Institute for Solid State Research,
             Heisenbergstrasse 1, D-70569 Stuttgart, Germany, }
\affiliation{\mbox{Marian Smoluchowski Institute of Physics, Jagiellonian University,
             Prof. S. \L{}ojasiewicza 11, PL-30348 Krak\'ow, Poland}}

\date{\today}

\begin{abstract}
Using the multiband $d-p$ model and unrestricted Hartree-Fock
approximation we investigate the electronic structure and spin-orbital
order in three-dimensional MnO$_3$ lattice such as realized in
LaMnO$_3$. The orbital order is induced and stabilized by particular
checkerboard pattern of oxygen distortions arising from the Jahn-Teller
effect in the presence of strong Coulomb interactions on $e_g$ orbitals
of Mn ions. We show that the spin-orbital order can be modeled using a
simple \textit{Ansatz} for local crystal fields alternating between two
sublattices on Mn ions, which have non-equivalent neighboring oxygen
distortions in $ab$ planes.
The simple and computationally very inexpensive $d-p$ model reproduces
correctly nontrivial spin-orbital order observed in undoped LaMnO$_3$.
Orbital order is very robust and is reduced by $\sim 3$ \% for large
self-doping in the metallic regime.
\end{abstract}

\maketitle

\section{Introduction}
\label{intro}

The spin and orbital ordering found in three dimensional LaMnO$_3$
perovskite is an old problem which is nowadays quite well understood
\cite{Dag01,Tok06},
see also early and recent experimental and theoretical references
\cite{ari93,sai95,mah95,miz95,saw97,cap00,tob01,rav02,end99,iva02,kat04,
eba05,zen05,gav06,pav10,kov10,yin06,yam06,kov11,gon11,mel15,gav17,jan18,sna18}.
The short summary and conclusions coming out from these papers are
as follows: At zero temperature the LaMnO$_3$ lattice has orthorhombic
symmetry. The lattice is distorted due to strong Jahn-Teller (JT)
effect: the MnO$_6$ octahedra are deformed, as shown schematically
using the simplified picture for the $ab$ ferromagnetic (FM) plane
(nonzero tilting of the octahedra is neglected in this study) in Fig. 1.
The magnetic moments on Mn ions correspond to spins $S\simeq 2$ and
the magnetic structure is of the $A$-type antiferromagnetic ($A$-AF),
i.e., FM order on separate $ab$ planes, coupled antiferromagnetically
plane-to-plane along the crystallographic $c$ axis. The electron
occupations on manganese $t_{2g}$ orbitals $\{xy,yz,zx\}$ are very
close to 1, while both $e_g$ orbitals, $\{x^2-y^2,3z^2-r^2\}$, contain
together roughly one (the fourth) electron. The orbital order which
settles in $e_g$ orbitals is not seen when using the standard
$\{x^2-y^2,3z^2-r^2\}$ basis
(which corresponds to choosing $z$ quantization axis direction).
However, if we consider a checkerboard pattern superimposed upon $ab$
plane and choose $x$ axis of quantization on ``black'' fields (MnO$_4$
rhombuses $m=2$ expanded along $x$ axis, see Fig. 1), and $y$ axis of
quantization on ``white'' fields (MnO$_4$ rhombuses $m=1$ expanded
along $y$ axis), then the orbital order becomes transparent: the
orbitals with majority electron-occupations follow the
\mbox{$3x^2-r^2$ / $3y^2-r^2$} pattern. Oxygen distortions repeat along
the $c$ axis, and $e_g$ orbitals follow. The common terminology to
describe this type of order is $C$-type alternating orbital ($C$-AO)
order.

\begin{figure}[t!]
\vskip -1.1cm
\begin{center}
\includegraphics[width=15.9cm]{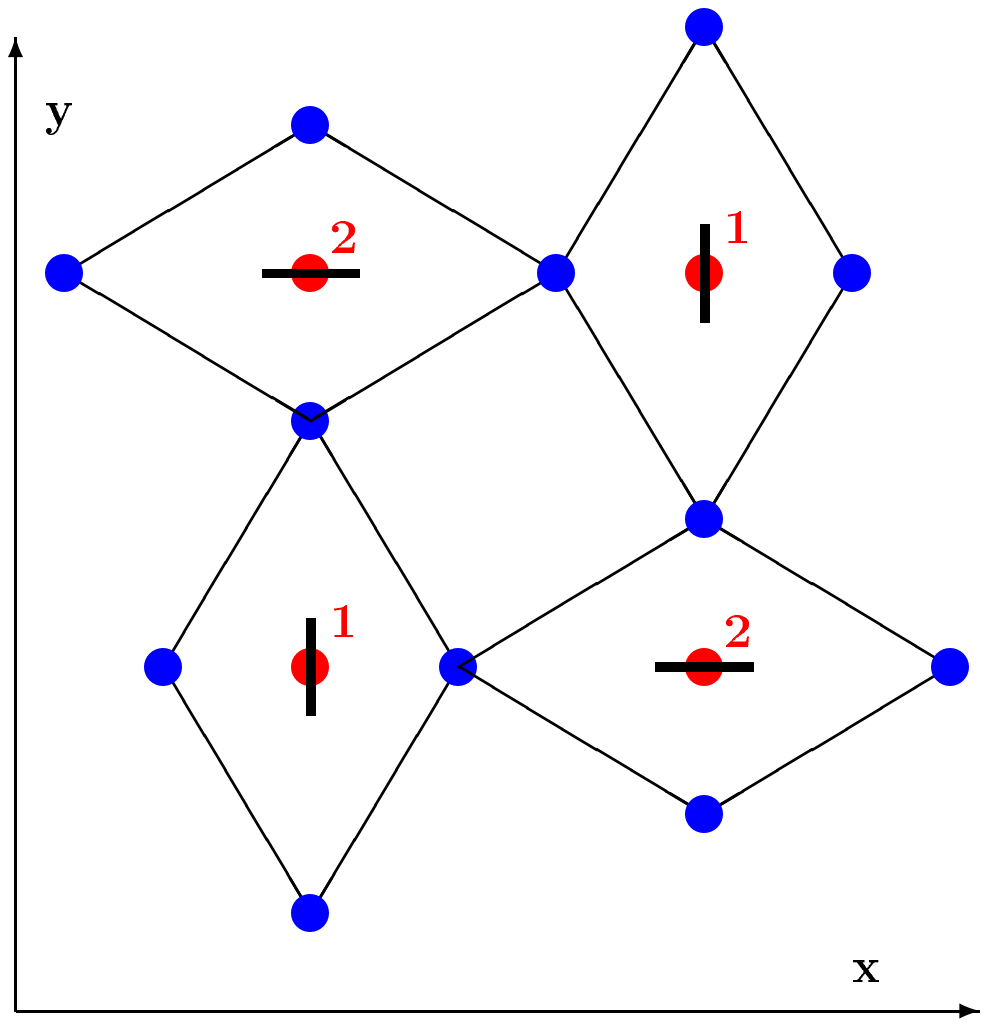}
\end{center}
\vspace{-12.8cm}
\caption{
Schematic view of JT distortions used for the Hartree-Fock (HF)
computations. A single $ab$ plane of the low-temperature phase of
LaMnO$_3$ is presented. Red/blue dots denote positions of
manganese/oxygen ions.
The long bars denote energy-privileged (due to local crystal fields)
$3y^2-r^2$ orbitals (at Mn$_1$) or $3x^2-r^2$ orbitals (at Mn$_2$ ions)
--- their cooperative arrangement corresponds to orbital order which
supports FM spin order (spins are not shown). The numbers close to
manganese positions identify different ions
(see the corresponding entries in Table II) as belonging to $3y^2-r^2$
($m=1$) and $3x^2-r^2$ ($m=2$) orbitals with lower energy. Note that
horizontal and vertical directions on the figure correspond to $x$ and
$y$ axes, respectively, which are at 45{\degree} angle to the
crystallographic $\{a,b\}$ axes. Orbital order is repeated in
consecutive layers along the $z$ (crystallographic $c$) axis.
}
\label{figJT}
\end{figure}

The origin of the orbital order was under debate using effective $3d$
models taking into account electron-electron and electron-lattice
interactions \cite{mil97,gri98,fei99,hot99,yin06}. It was established
that large JT splitting and superexchange are together responsible for
$C$-AO order observed in LaMnO$_3$ \cite{gri98,fei99,hot99}.
Surprisingly, electron-lattice interaction is too weak to induce the
orbital order alone and has to be helped by electron-electron
superexchange \cite{yin06}. Importance of the joint effect of strong
correlations and Jahn-Teler distortions was concluded from the dynamical
mean field theory \cite{yam06}. On the other hand, superexchange alone
would stabilize $C$-AO order but the order would be fragile and the
orbital ordering temperature would be too low \cite{fei99,ole05}.

The large $S=2$ spins at Mn$^{3+}$ ions are coupled by spin-orbital
superexchange which stabilizes $A$-AF spin order in LaMnO$_3$ below
N\'eel temperature $T_{\rm N}=140$ K. However, spin order parameter
could be little reduced if the {\it ideal ionic model} approximation
does not strictly apply and the actual number of electrons transferred
from La to MnO$_3$ subsystem is reduced to $3-x$, i.e., one has
La$^{+(3-x)}$(MnO$_3$)$^{-3+x}$,
where $x$ is so-called \textit{self-doping}.

The idea of self-doping comes directly from \textit{ab-initio}
computations and may be addressed when $d-p$ hybridization is
explicitly included in the $d-p$ model. There, a rather
trivial know-how is that charges on cations, computed using simple
Mulliken population analysis --- or better Bader population analysis
--- are almost never the same as in the idealized ionic-model.
Therefore, it is quite reasonable to introduce into the $d-p$ model the
idea that each La$^{+(3-x)}$ donates onto MnO$_3$ lattice on average
not 3 but $3-x$ electrons. The self-doping can be regarded as a free
parameter which could be adjusted using some additional experimental
data. However, on the other hand, one can compute $x$ using
\textit{ab-initio} or density functional theory (DFT) with local
Coulomb interaction $U$ (DFT+$U$) cluster computations, exactly like
it was done in Ref. \cite{Ros17}. This is, however, very costly,
and contradicts one of the most important virtues of the $d-p$ model,
i.e., time and cost efficiency of the computations.

Returning now to $x$ in LaMnO$_3$, we note that if $x$ is finite but
still small enough, other magnetic phases could be more stable --- in
particular ordinary FM order is found frequently. Such a situation was
studied experimentally in Sr$_x$La$_{1-x}$MnO$_3$ \cite{iva02}; note
that here subscript $x$ in the chemical formula Sr$_x$La$_{1-x}$MnO$_3$
which routinely is identified as ordinary {\it doping} is only roughly
similar to our {\it self-doping}~$x$. When we  increase ordinary doping
$x$ from very small towards intermediate values, then the metallic
regime sets in Sr$_x$La$_{1-x}$MnO$_3$ \cite{Tok06}.

The purpose of this short paper is to investigate whether the
spin-orbital order observed in LaMnO$_3$ would arise in the multiband
$d-p$ model with the explicit treatment of oxygen $2p$ electrons. Here
we focus on simple modeling of the JT effect, where oxygen distortions
are treated as semiempirical input for the electronic $d-p$ model.
To make the model realistic, we also include in it non-zero Coulomb
repulsion on oxygen ions and spin-orbit interaction on Mn ions. Note
that model studies done on LaMnO$_3$ before usually neglected these
Hamiltonian components. In addition, we would like to account for the
possibility of non-trivial self-doping $x\neq 0$
(just as found before in ruthenium, iridium, titanium, and vanadium
oxides \cite{Ros15,Ros16,Ros17,Ros18}).

The paper is organized as follows. We introduce the model Hamiltonian
and its parameters in Sec. \ref{sec:model}. The numerical method is
presented in Sec. \ref{sec:num}, where we treat the JT effect in a
semi-empirical way (Sec. \ref{sec:jt}), and present the unrestricted
Hartree-Fock approximation in Sec. \ref{sec:hf} as a method of choice
to describe the states with broken spin-orbital symmetries. It was
shown recently that this approach gives very reliable results for
doped vanadium perovskites \cite{Ave18}.
The magnetization direction is there one of the open questions and we
discuss possible states in Sec. \ref{sec:sz}. The ground state of
LaMnO$_3$ is described in Secs. \ref{sec:x=0}, \ref{sec:x=1/16}, and
\ref{sec:x=1/8} for three self-doping levels, $x=0$, $x=1/16$, and
$x=1/8$. Finally, we also remark on the consequences of neglected $e_g$
and $t_{2g}$ splittings in Sec. \ref{sec:nos}. In Sec. \ref{sec:oo} we
present the dependence of orbital order parameter on the self-doping
level and conclude that the order is robust.
A~short summary and conclusions are presented in Sec. \ref{sec:summa}.

\section{Hamiltonian}
\label{sec:model}

We introduce the multiband $d-p$ Hamiltonian for MnO$_3$
three-dimensional $4\times 4\times 4$ (periodic boundary conditions)
cluster which includes five $3d$ orbitals at each manganese ion and
three $2p$ orbitals at each oxygen ion,
\begin{equation}
{\cal H}=
H_{dp}+H_{pp}+H_{\rm so}+H_{\rm diag}+H_{\rm int}^d+H_{\rm int}^p.
\label{model}
\end{equation}
where $H_{dp}$ stands for the $d-p$ hybridization and $H_{pp}$ for the
interoxygen $p-p$ hopping, $H_{\rm so}$ is the spin-orbit coupling,
$H_{\rm diag}$ is the diagonal part of kinetic energy
(bare levels energies and local crystal fields). Here
$H_{\rm int}^d$ and $H_{\rm int}^p$ stand for the intraatomic Coulomb
interactions at Mn and O ions, respectively.
Optionally one could add JT energy as $H_{\rm JT}$.
However, instead of introducing this term with a quite complicated form
\cite{mul10}, we shall model the JT interaction by a simple
\textit{Ansatz} which can be inserted into $H_{\rm diag}$ as local
potentials acting on $e_g$ electrons.
The cluster geometry and precise forms of different terms are standard;
these terms were introduced in the previous realizations of the $d-p$
model devoted to transition metal oxides \cite{Ros15,Ros16,Ros17,Ros18}.

The kinetic energy in the Hamiltonian (\ref{model}) consists of:
\begin{eqnarray}
H_{dp}&=&\sum_{\{m\alpha;j\nu\},\sigma}\left(t_{m\alpha;j\nu}
 d^{\dagger}_{m\alpha,\sigma}p_{j\nu,\sigma}^{} + {\rm H.c.}\right),\\
H_{pp}&=&\sum_{\{i\mu;j\nu\},\sigma}\left(t_{i\mu;j\nu}
 p^{\dagger}_{i\mu,\sigma}p_{j\nu,\sigma}^{} + {\rm H.c.}\right),
\end{eqnarray}
where $d_{m\alpha,\sigma}^{\dagger}$ ($p_{j\nu,\sigma}^{\dagger}$)
is the creation operator of an electron at manganese site $m$
(oxygen site $i$) in an orbital $\alpha$ ($\nu$) with up or down spin,
$\sigma=\uparrow,\downarrow$. The model includes all five $3d$ orbital
states $\alpha\in\{xy,yz,zx,x^2-y^2,3z^2-r^2\}$, and three $2p$ oxygen
orbital states, $\{\mu,\nu\}\in\{p_x,p_y,p_z\}$.
In the following we will use shorthand notation, and instead of
$\{x^2-y^2,3z^2-r^2\}$ we shall write $\{(\bar{z}),(z)\}$ ---
this emphasizes the fact that $z$ axis is chosen as the quantization
axis for this $e_g$ orbital basis, while () brackets are here to
distinguish these Mn($3d$) orbitals from O($2p$) $\{x,y,z\}$ orbitals.
The matrices $\{t_{m\alpha;j\nu}\}$ and $\{t_{i\mu;j\nu}\}$ are assumed
to be non-zero only for nearest neighbor manganese--oxygen $d-p$ pairs,
and for nearest neighbor oxygen--oxygen $p-p$ pairs. The next nearest
neighbor hoppings are neglected. (The nonzero $\{t_{m\alpha;j\nu}\}$
and $\{t_{i\mu;j\nu}\}$ elements are listed in the Appendix of
Ref.~\cite{Ros15}).

The spin-orbit part,
$H_{\rm so}=\zeta\sum_i\textbf{L}_i\cdot\textbf{S}_i$,
is a one-particle operator (scalar product of angular momentum and spin
operators at site $i$), and it can be represented in the form similar
to the kinetic energy $H_{\rm kin}$ \cite{miz96,Pol12,Mat13,Du13},
\begin{equation}
H_{\rm so}= \sum_m\left\{\sum_{\alpha\neq\beta;\sigma,\sigma'}
t^{so}_{\alpha,\sigma;\beta,\sigma'}
d^{\dagger}_{m\alpha,\sigma}d_{m\beta,\sigma'}^{}+\mathrm{H.c.}\right\},
\label{so-part}
\end{equation}
with $t^{so}_{\alpha,\sigma;\beta,\sigma'}$ elements restricted to
single manganese sites. They all depend on spin-orbit coupling strength
$\zeta$ ($\zeta\approx 0.04$ eV \cite{Dai08}), which is weak but still
it influences the preferred spin direction in the $A$-AF phase.
For detailed formulas and tables listing
$\{t^{so}_{\alpha,\sigma';\beta,\sigma}\}$ elements,
see Refs. \cite{Ros15,Pol12}.

The diagonal part $H_{\rm diag}$ depends on electron number operators.
It takes into account the effects of local crystal fields and the
difference of reference orbital energies
(here we employ the electron notation),
\begin{equation}
\Delta=\varepsilon_d-\varepsilon_p,
\label{Delta}
\end{equation}
between $d$ and $p$ orbitals (for bare orbital energies) where
$\varepsilon_d$ is the average energy of all $3d$ orbitals, i.e., the
reference energy before they split in the crystal field. We fix this
reference energy for $d$ orbitals to zero, $\varepsilon_d=0$, and use
only $\Delta=-\varepsilon_p$ and the crystal field splittings
$\{f_{\mu,\sigma}^{cr}\}$ as parameters, thus we write
\begin{eqnarray}
H_{\rm diag}=
\sum_{i;\mu=x,y,z;\sigma}
\varepsilon_p^{} p^\dagger_{i\mu,\sigma}p_{i\mu,\sigma}^{} \nonumber \\
+\sum_{m;\alpha=xy,yz,... ;\sigma}
f^{cr}_{\mu,\sigma} d^\dagger_{m\alpha,\sigma}d_{m\alpha,\sigma}^{}.
\end{eqnarray}
The first sum is restricted to oxygen sites $\{i\}$, while the second
one runs over manganese sites $\{m\}$. The crystal-field splitting
strength vector ($f_{\alpha,\sigma}^{cr}$) describes the splitting
within $t_{2g}$ and within $e_g$ levels, as well as the $t_{2g}$ to
$e_g$ splitting, respectively.

\begin{figure}[t!]
\vspace{-3.6cm}
\begin{center}
\includegraphics[width=15.2cm]{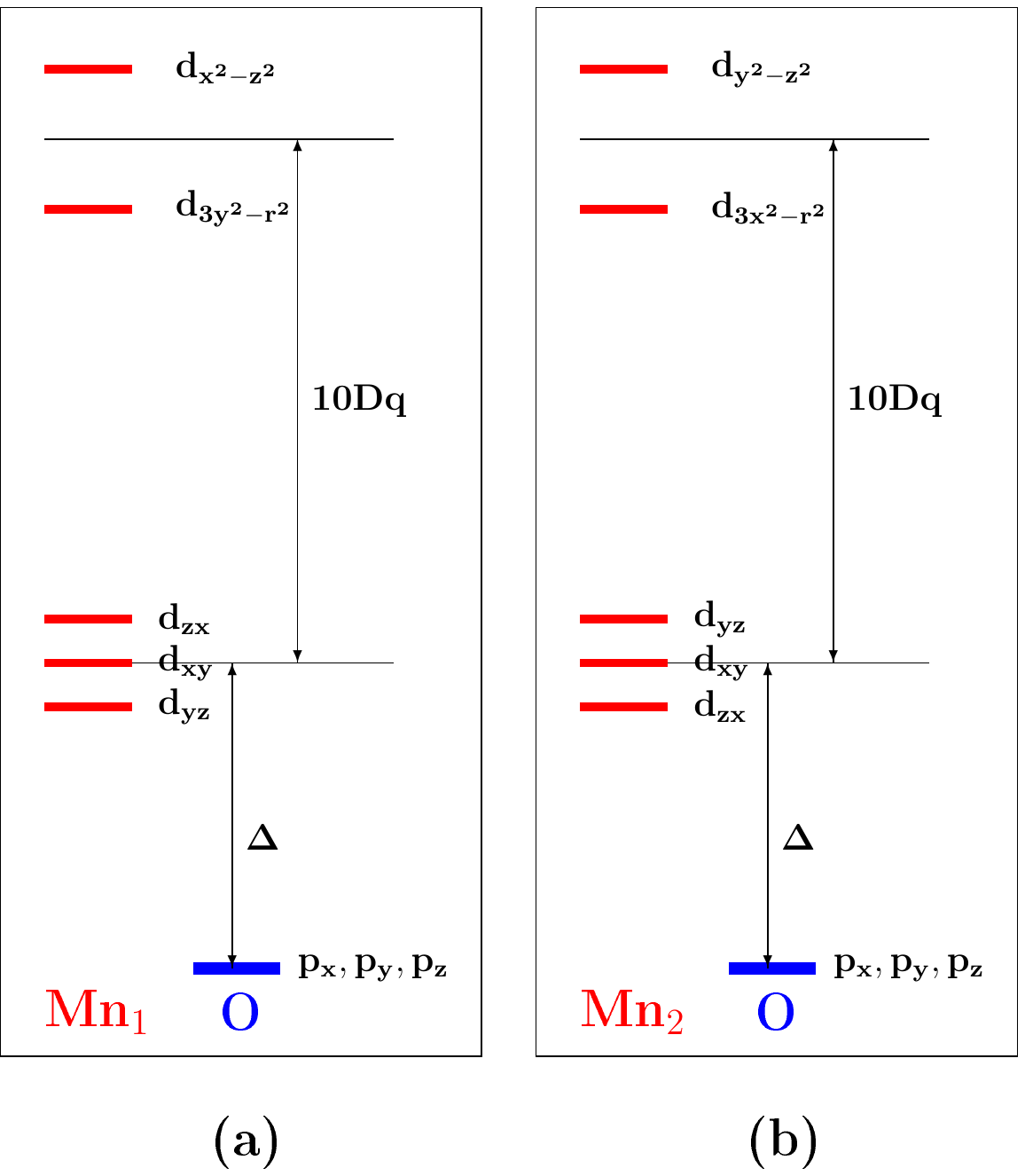}
\end{center}
\vspace{-8.1cm}
\caption{%
Artist's view of the bare $3d$ levels (no interaction) at Mn$_m$
($m=1,2$) ions split by local crystal fields originating from JT
distortions, $2p$ oxygen levels in the $d-p$ model.
(a) Left panel: $yz$ orbital is lower than $zx$ orbital which
corresponds to O$_4$ square when distorted from an ideal square into
rhombus elongated along $y$ direction (see Mn$_1$ ions in Fig.~1),
$e_g$ levels are also split --- $3y^2-r^2$ is lower than $x^2-z^2$
(here $y$-axis is chosen as the quantization axis).
(b) Right panel: $zx$ orbital is lower than $yz$ orbital which
corresponds to O$_4$ square when distorted from an ideal square into
rhombus elongated along $x$ direction (see Mn$_2$ ions in Fig. 1),
$e_g$ levels are also split --- $3x^2-r^2$ is lower than $y^2-z^2$
(here $x$-axis is chosen as the quantization axis).
The average distance between $t_{2g}$ levels and $e_g$ levels ($10Dq$)
is $\sim 1.7$ eV \cite{zam02}, and $\Delta$ is oxygen-to-manganese
charge-transfer energy.
}
\label{figJT}
\end{figure}

Furthermore, the distance $10Dq$ between $t_{2g}$ levels and $e_g$
levels is $\sim 1.7$ eV \cite{zam02}). For the group of $t_{2g}$ levels
and their splittings in accordance with local JT distortion of
particular MnO$_6$ octahedron \textit{we assume} that either $yz$ is
lower than $zx$ orbital which should correspond to O$_4$ square
(in $ab$ plane) when distorted from an ideal square into rhombus
elongated along $y$ direction, or the opposite: $zx$ is lower than $yz$
orbital which should correspond to O$_4$ distorted into rhombus
elongated along $x$ direction (compare Fig.~2). The splitting value
should be $\sim 0.1$ eV what is an educated guess
(compare Ref. \cite{Ros18}) or even smaller \cite{gav06}. What concerns
the occupied $e_g$ levels --- they are also split. We assume varying
(ion to ion) local crystal fields and choose appropriately the local
axes of quantization, following Mn sublattices shown in Fig. 1.

In rhombuses elongated along the $x$ axis (see Fig.~1), we use $3d$
orbital states at Mn sites $\{m\}$,
\mbox{$\{\alpha,\beta\}\in\{xy,yz,zx,y^2-z^2,3x^2-r^2\}$} and
correspondingly new $\{D_{m\alpha,\sigma}^{\dagger}\}$ creations
operators; the new form of local crystal field is
\mbox{$f_{\mu,\sigma} D^\dagger_{m\mu,\sigma}D_{m\mu,\sigma}$}.
Here our shorthand notation for $\{y^2-z^2,3x^2-r^2\}$ orbitals will be
$\{(\bar{x}),(x)\}$ (now $x$ is chosen as the quantization axis).
On the other hand, in rhombuses elongated along the $y$ axis
(see Fig. 1),  we use the following
$3d$ orbital states \mbox{$\alpha\in\{xy,yz,zx,x^2-z^2,3y^2-r^2\}$} and
correspondingly new $\tilde{D}_{j,\nu,\sigma}^{\dagger}$ creation
operators and the new form of local crystal field
$f_{\alpha\sigma}^{cr}\tilde{D}^\dagger_{m\alpha,\sigma}\tilde{D}_{m\beta,\sigma}^{}$.
Here our shorthand notation for $\{z^2-x^2,3y^2-r^2\}$ orbitals is
$\{(\bar{y}),(y)\}$ (now $y$ is chosen as the quantization axis).

Returning now to rhombuses expanded along the $x$ axis when we work
with $D_{(x)\sigma}^{\dagger}$ and $D_{(\bar{x})\sigma}^{\dagger}$
operators ($x$ quantization axis) the corresponding canonical
transformation is:
\begin{equation}
\left( \begin{array}{c}   {D}_{(\bar{x}) \sigma}^{\dagger}  \\
{D}_{(x) \sigma}^{\dagger} \end{array} \right) =
\left( \begin{array}{cc}  -\frac{1}{2} &  - \frac{\sqrt{3}}{2} \\
    \frac{\sqrt{3}}{2}   &   - \frac{1}{2} \end{array} \right)
\left( \begin{array}{c}   {d}_{(\bar{z}) \sigma}^{\dagger}  \\
{d}_{(z) \sigma}^{\dagger} \end{array} \right),
\label{DD}
\end{equation}
where $d$ operators are standard (i.e., for the $z$ quantization axis).

We can also compute the  operators of particle numbers, which are:
\begin{eqnarray}
\label{Dxbar}
D_{(\bar{x})\sigma}^{\dagger} D_{(\bar{x})\sigma}^{} &=&
  \frac{1}{4} d_{(\bar{z})\sigma}^{\dagger}  d_{(\bar{z})\sigma}^{} +
  \frac{3}{4}   d_{(z)\sigma}^{\dagger}  d_{(z)\sigma}^{} \nonumber \\
&+& \frac{\sqrt{3}}{4}\left(  d_{(\bar{z})\sigma}^{\dagger}d_{(z)\sigma}^{}
+  d_{(z)\sigma}^{\dagger}  d_{(\bar{z})\sigma}^{} \right),  \\
\label{Dx}
D_{(x)\sigma}^{\dagger} {D}_{(x)\sigma}^{} &=&
\frac{3}{4} d_{(\bar{z})\sigma}^{\dagger}  d_{(\bar{z})\sigma}^{} +
       \frac{1}{4}   d_{(z)\sigma}^{\dagger}  d_{(z)\sigma}^{} \nonumber \\
&-& \frac{\sqrt{3}}{4}\left(  d_{(\bar{z})\sigma}^{\dagger}d_{(z)\sigma}^{}
+  d_{(z)\sigma}^{\dagger}  d_{(\bar{z})\sigma}^{} \right).
\end{eqnarray}
Note that orbitals $t_{2g}$ remain the same as before, i.e.,
they were not transformed.

When looking at the formulae just above we immediately see that while
in $H_{\rm diag}$ the part with local crystal field is formally
expressed by using
$f^{cr}_{\mu,\sigma} D^\dagger_{m,\mu,\sigma}D_{m,\mu,\sigma}^{}$,
then in fact, thanks to Eqs. (\ref{Dxbar}) and (\ref{Dx}) we can still
work with old standard $d$ operators ($z$ quantization axis).

Returning to rhombuses expanded along the $y$ axis, we should work with
$\tilde{D}_{(\bar{y})\sigma}^{\dagger}$ and
$\tilde{D}_{(y)\sigma}^{\dagger}$ operators (for $y$ quantization axis)
and the corresponding formulae are:
\begin{equation}
\left( \begin{array}{c}   \ \tilde{D}_{(\bar{y}) \sigma}^{\dagger} \\
\tilde{D}_{(y) \sigma}^{\dagger} \end{array} \right) =
\left( \begin{array}{cc}   \frac{1}{2} &  - \frac{\sqrt{3}}{2} \\
    -\frac{\sqrt{3}}{2}   &   - \frac{1}{2}      \end{array} \right)
\left( \begin{array}{c}   {d}_{(\bar{z}) \sigma}^{\dagger}  \\
{d}_{(z) \sigma}^{\dagger} \end{array} \right)
\end{equation}
The particle number operators are:
\begin{eqnarray}
\tilde{D}_{(\bar{y})\sigma}^{\dagger} \tilde{D}_{(\bar{y})\sigma}^{} &=&
  \frac{1}{4} d_{(\bar{z})\sigma}^{\dagger}  d_{(\bar{z})\sigma}^{}
  +\frac{3}{4}   d_{(z)\sigma}^{\dagger}  d_{(z)\sigma}^{} \nonumber \\
&-&\frac{\sqrt{3}}{4}\left(d_{(\bar{z})\sigma}^{\dagger}d_{(z)\sigma}^{}
+  d_{(z)\sigma}^{\dagger}  d_{(\bar{z})\sigma}^{} \right),  \\
\tilde{D}_{(y)\sigma}^{\dagger} \tilde{D}_{(y)\sigma}^{} &=&
  \frac{3}{4} d_{(\bar{z})\sigma}^{\dagger}  d_{(\bar{z})\sigma}^{}
  +\frac{1}{4}   d_{(z)\sigma}^{\dagger}  d_{(z)\sigma}^{} \nonumber \\
&+&\frac{\sqrt{3}}{4}\left(d_{(\bar{z})\sigma}^{\dagger}d_{(z)\sigma}^{}
  +  d_{(z)\sigma}^{\dagger}  d_{(\bar{z})\sigma}^{} \right).
\end{eqnarray}

To make a short summary: the subsequent HF computations will be
performed in the framework of the standard basis (i.e., old basis with
$z$ quantization axis) and after reaching self-consistency, we extract
occupation numbers using formulae from Eqs. (\ref{Dxbar}) and (\ref{Dx});
to give an example the electron occupation in $3x^2-r^2$ orbitals is just
\begin{eqnarray}
\!\left\langle {D}_{(x)\sigma}^{\dagger}{D}_{(x)\sigma}^{}\right\rangle&=&
\frac{3}{4}\left\langle d_{(\bar{z})\sigma}^{\dagger}d_{(\bar{z})\sigma}^{}\right\rangle
+\frac{1}{4}\left\langle d_{(z)\sigma}^{\dagger}d_{(z)\sigma}^{}\right\rangle
\nonumber  \\
&-&\frac{\sqrt{3}}{4}\left\langle d_{(\bar{z})\sigma}^{\dagger}d_{(z)\sigma}^{}
+ d_{(z)\sigma}^{\dagger}d_{(\bar{z})\sigma}^{} \right\rangle,
\end{eqnarray}
\textit{etcetera}.

The on-site Coulomb interactions $H_{\rm int}^d$ for $d$ orbitals take
the form of a degenerate Hubbard model \cite{Ole83}
\begin{eqnarray}
H_{\rm int}^d&=&
\sum_{m,\alpha<\beta}\left(U_d-\frac{5}{2} J^d_{\alpha\beta}\right)
n_{m\alpha} n_{m\beta}\nonumber\\
&+&U_d \sum_{m\alpha} n_{m\alpha,\uparrow} n_{m\alpha,\downarrow}
-2\sum_{m,\alpha<\beta}J^d_{\alpha\beta}\,\vec{S}_{m\alpha}\cdot\vec{S}_{m\beta}
\nonumber\\
&+&  \sum_{m,\alpha\neq\beta} J^d_{\alpha\beta}\,
d^\dagger_{m\alpha,\uparrow} d^\dagger_{m\alpha,\downarrow}
d_{m\beta,\downarrow}^{}d_{m\beta,\uparrow}^{},
\label{hubbard2-intra}
\end{eqnarray}
where $n_{m\alpha}=\sum_{\sigma}n_{m\alpha,\sigma}$ is the electron
density operator in orbital $\alpha$; $\{\alpha,\beta\}$ enumerate
different $d$ orbitals, and $J_{d,\alpha\beta}$ is the tensor of
on-site interorbital exchange (Hund's) elements for $d$ orbitals;
$J^d_{\alpha\beta}$ has different entries for the $\{\alpha,\beta\}$
pairs corresponding to two $t_{2g}$ orbitals ($J_{\rm H}^t$), and for a
pair of two $e_g$ orbitals ($J_{\rm H}^e$), and still different for the
case of cross-symmetry terms \cite{ole05,Hor07}; all these elements are
included and we assume the Racah parameters: \mbox{$B=0.1$ eV} and
$C=4B$.

The local Coulomb interactions $H_{\rm int}^p$
at oxygen sites (for $2p$ orbitals) are analogous,
\begin{eqnarray}
H_{\rm int}^p&=&
\sum_{i,\mu<\nu,\sigma} \left(U_p-\frac{5}{2} J^p_{\rm H}\right)
n_{i\mu} n_{i\nu} \nonumber\\
&+&U_p \sum_{i\mu}  n_{i\mu,\uparrow} n_{i\mu,\downarrow}
-2J^p_{\rm H}\sum_{i,\mu<\nu} \vec{S}_{i\mu}\cdot\vec{S}_{i\nu}
\nonumber\\
&+& J^p_{\rm H}\sum_{i,\mu\neq\nu}
p^\dagger_{i\mu\uparrow}p^\dagger_{i\mu\downarrow}
p_{i\nu\downarrow}^{}p_{i\nu\uparrow}^{},
\label{hubbard3-intra}
\end{eqnarray}
where the intraatomic Coulomb repulsion is denoted as $U_p$ and all
off-diagonal elements of the tensor $J^p_{\mu\nu}$ are equal
(as they connect the orbitals of the same symmetry), i.e.,
$J^p_{\mu\nu}\equiv J^p_{\rm H}$. Up to now, interaction at oxygen
ions $H_{\rm int}^p$ were neglected in the majority of studies
(i.e., for simplicity it was being assumed that $U_p=J^p_{\rm H}=0$).

\begin{table}[b!]
\caption{Parameters of the multiband model (\ref{model}) (all in eV)
used in HF calculations. For the hopping integrals we adopt the values
from Refs. \cite{miz95,miz96} and include oxygen-oxygen hopping
elements in $H_{pp}$, given by $(pp\sigma)=0.6$, $(pp\pi)=-0.15$ eV
(here we use the Slater notation \cite{Sla54}). The charge-transfer
energy $\Delta$ (\ref{Delta}) is defined for bare levels.
The magnitude of splitting within $t_{2g}$ and $e_g$ levels
is arbitrarily taken as 0.1 eV and 0.2 eV.
}
\begin{ruledtabular}
\begin{tabular}{ccccccccc}
$\zeta $ & $(pd\sigma)$ & $(pd\pi)$ & $\Delta$
                & $U_d$ & $J_{\rm H}^t$ & $J_{\rm H}^e$
                & $U_p$ & $J_{\rm H}^p$   \\
  \hline
0.04 & $-1.8$ & 0.9 & 2.0 & 8.0 & 0.8 & 0.9 & 4.4 & 0.8   \\
\end{tabular}
\end{ruledtabular}
\label{tab:para}
\end{table}

In the following we use the parameters $U_d$, $J^d_{\mu\nu}$, $U_p$,
and $J_{\rm H}^p$ similar to those used before for titanium and
vanadium oxides \cite{Ros16,Ros17,Ros18};
for the hopping integrals we follow the work by Mizokawa and Fujimori
\cite{miz95,miz96}. The value $U_p\sim 4.0$ eV was previously used in
copper oxides \cite{Hybe92,Sing13}. Concerning the parameter $\Delta$
an educated guess is necessary. Old-fashioned computations, such as
those reported in the classical textbook of Harrison \cite{Har05} and
shown in tables therein suggest 1.5 eV. Results of a more detailed
study suggest that $\Delta<2.5$ eV \cite{Rau05}. We also use the
Slater notation \cite{Sla54}.
We performed our computations for the parameter set in \mbox{Table I}.

Our reference system is LaMnO$_3$ where the total electron number in
the $d-p$ subsystem is $N_e=18+4=22$ per one MnO$_3$ unit provided we
assume an ideal ionic model with no self-doping ($x=0$), i.e., all
three La valence electrons are transferred to MnO$_3$ unit. Another
possibility is to take finite but small self-doping $x$. The problem
how to fix $x$ is a difficult question. The best way is to perform
independent, auxiliary \textit{ab-initio} or local density approximation
(LDA) with Coulomb interaction $U$ (LDA+$U$) computations and to
extract the electronic population on the cation $R$ (in $R$MnO$_3$)
analogously like it was done in Ref. \cite{Ros17}. This is however
quite expensive. Without such auxiliary \textit{ab-initio} computations
one is left either with speculations or one should perform computations
using several different values of $x$.

\section{Numerical studies}
\label{sec:num}

\subsection{Computational problems concerning the Jahn-Teller Hamiltonian}
\label{sec:jt}

Now let us return once more to the important part of the electronic
Hamiltonian in perovskites, namely to the influence of JT distortions
on the electronic structure. These rarely can be treated in a direct
(and exact) way during the computations. Most often a semiempirical
treatment of JT terms is used: namely one assumes an explicit form and
the magnitudes of the lattice distortions, such as suggested by
experiments. Thus, the distorted lattice is frozen and we take this as
the experimental fact. The JT modes
and the JT Hamiltonian do not enter computations anymore --- their
only role is to deform the lattice and to change Mn$-$O distances.
Instead, one collects all Mn$-$O and O$-$O bond lengths (as suggested
by experiment) and because of modified bond lengths one modifies the
matrix of kinetic hopping parameters. In this respect quite popular is
the Harrison scaling \cite{Har05}. We have used it in the present study.

The second important consequence of changed Mn$-$O distances is the
creation of additional local crystal fields  (in addition to standard
crystal field  which is responsible for $t_{2g}$ to $e_g$ splitting).
These additional local crystal fields renormalise bare energy levels
within $e_g$ doublets and within $t_{2g}$ triplets.
(Note that this picture is valid at the level of static one-particle
effective-potential approximation;
i.e., it is similar like crystal-field cubic potential approximation
which gives rise to standard $10Dq$ splitting).
Thus as the second part of modeling JT effect, what we do is:
$e_g$ doublets and $t_{2g}$ triplets will be split as already discussed
(in the previous Section) for $H_{\rm diag}$ and $f^{cr}_{\mu,\sigma}$.

\subsection{Unrestricted Hartree-Fock computations}
\label{sec:hf}

We use the unrestricted HF approximation (UHF) (with a single
determinant wave function) to investigate the $d-p$ model (\ref{model}).
The technical implementation is the same as that described in Refs.
\cite{miz95,miz96,Ros15,Ros16} featuring the averages
$\langle d^\dagger_{m\alpha,\uparrow}d_{m\alpha,\uparrow}^{}\rangle$
and $\langle p^\dagger_{i\mu,\uparrow} p_{i\mu,\uparrow}^{}\rangle$
(in the HF Hamiltonian) which are treated as order parameters. We use
the $4\times 4\times 4$ clusters which are sufficient for the present
$d-p$ model with only nearest neighbor hopping terms. During
HF iterations the order parameters are recalculated self-consistently
until convergence. The studied scenarios for the ground state symmetry
were those with spin order: FM, $A$-AF, $G$-AF (N\'eel state), $C$-AF
(AF in $ab$ plane, repeated in the consecutive $ab$ planes when moving
along the $c$ axis), or non-magnetic; the considered easy magnetization
direction was either $x$ or $z$.

To improve HF-convergence we used the quantum chemistry technique
called level shifting \cite{Sou73}. It is based on replacing the true
HF Hamiltonian by a different Hamiltonian --- the one with the
identical eigenvectors (one particle eigenfunctions) as the original
Hamiltonian and with identical \textit{occupied} eigenenergies. The
original eigenenergies of virtual states are however uniformly shifted
upwards by a fixed constant value. When applying virtual level shifting
we can obtain some additional information. Namely when the splitting
between the highest occupied molecular orbital (HOMO) and the lowest
unoccupied molecular orbital (LUMO), i.e., HOMO-LUMO splitting (after
correcting for the shift) is negative or zero, then the
single-determinant HF ground state we obtained is not correct and
usually the reason is that the true ground state is in fact conducting.
We remind that the HOMO-LUMO gap serves here as an estimate of the
experimental band gap.

\subsection{Searching for the magnetization direction}
\label{sec:sz}

We performed computations for several values of self-doping $x$.
They give the orbital order for $e_g$ orbitals of $C$-type alternating
orbital ($C$-AO) order in the regime of low self-doping $x<\frac{1}{8}$
(see Table II). At the same time, the spin order is $A$-AF, with the
easy-axis of magnetization along the $x$ direction. The preferred spin
direction is however not generic as the ground states with $z$-easy
axis and with $x$-easy axis are almost degenerate (within accuracy
below 1 meV). Average spin values on Mn ions are very close to $S=2$
and the HOMO-LUMO gaps $G$ were also computed, see Table II.

Finally, we remark on the magnetic state obtained in HF calculations:
The up- and down-spin occupations are equal, i.e.,
$\langle d^\dagger_{m\alpha,\uparrow}d_{m\alpha,\uparrow}^{}\rangle=
\langle d^\dagger_{m\alpha,\downarrow}d_{m\alpha,\downarrow}^{}\rangle$,
thus the average $z$-th spin component vanishes. However, this does
not imply that the found ground states are nonmagnetic. We have found
that the symmetry breaking with magnetization along $x$ or $y$ axis is
equivalent, and the averages of the type,
$\langle d^\dagger_{m\alpha,\uparrow}d_{m\alpha,\downarrow}^{}\rangle$,
are finite (not shown, but it is always the case for the data in
Section \ref{sec:gs}). When the summation over $\mu$ is performed,
i.e., if we calculate $Re\big\{\sum_\alpha\langle
d^\dagger_{m\alpha,\uparrow}d_{m\alpha,\downarrow}^{}\rangle\big\}$,
we obtain the average spin component along the $x$ direction,
$|\langle S^x\rangle|$. This provides evidence that the spins are
indeed aligned along the $x$ axis, and we give the average magnetization
$|\langle S^x\rangle|$ in Sec. \ref{sec:oo}. The imaginary part of the
same sum (if finite) does correspond to the average spin component along
the $y$ direction.

\section{Ground state of L\lowercase{a}M\lowercase{n}O$_3$}
\label{sec:gs}

\subsection{Zero self-doping $x=0$}
\label{sec:x=0}

The ground state of LaMnO$_3$ has $C$-AO orbital order, and this is
reproduced by the UHF calculations, see Table II, densities given in
boldface. Note that when using only the standard orbital basis
(i.e., the orbitals corresponding to the $z$ quantization axis), the
orbital order is completely hidden. To get more insightful results and
to describe the orbital order induced by lattice distortions, we
considered all the types of O$_4$ rhombuses.

It is important to realize that orbital order may be easily detected
only for properly selected orbitals, depending on the sublattice.
First, if the standard basis $\{3z^2-r^2,x^2-y^2\}$ is used, no trace
of any orbital order is seen, see Table II. For the other two possible
$e_g$ bases, $\{3x^2-r^2,y^2-z^2\}$ and $\{3y^2-r^2,z^2-y^2\}$, one
finds that the directional orbital has large electron density only on
one sublattice, but on the other sublattice this is not the case.
In other words, if one selects the $x$ quantization axis, the orbital
order is easily visible through a distinct asymmetry between Mn ions
at this and the other sublattice, i.e., in positions $m=1$ and $m=2$,
see Fig. 1. Thus, the found asymmetry in the density distribution
indicates that the order is of checkerboard type. Indeed, the
checkerboard pattern of oxygen distortions requires choosing different
local bases at two sublattices: on one with $x$ quantization axis, and
with $y$ quantization axis on the other. Then the orbital order is
clearly visible and the symmetry is correctly recovered, see the
electron densities listed in boldface in Table~II.

The spin order coexisting with $C$-AO order is $A$-AF, with the easy
axis of magnetization along the $x$ ($y$) direction, see Table II. Note
that the state with the same characteristics but with easy magnetization
axis along the $z$ direction (not shown in Table II) is only by 0.3 meV
higher (thus these two states are almost degenerate). The other HF
states with $C$-AO order and with ordinary FM spin order are by 2 meV
higher, while the states with $G$-AF or $C$-AF spin order are by 3.5 meV
higher than the ground state. Nonmagnetic state is never realized.

\begin{table}[t!]
\caption{Spin-orbital order and electron densities
$\langle n_{m\alpha}\rangle$ obtained on non-equivalent Mn ions for the
HF ground state (at zero temperature) as obtained for orthorhombic
LaMnO$_3$ at self-doping $x=0$ and $x=0.0625$.
The index $m=1,2$ denotes a Mn site of a given sublattice, as shown in
Fig. 1. Numbers in bold indicate the most appropriate quantization
direction, i.e., the best local orbital basis for the description of
orbital order at a given sublattice.
The HF calculations are summarized by: the HF energy per MnO$_3$ unit
cell, $E_{\rm HF}$, and the HOMO-LUMO gap $G$, and
the average magnetization value.
}
\begin{ruledtabular}
  \begin{tabular}{ccccc}
self-doping & \multicolumn{4}{c}{\textbf{$A$-AF / $C$-AO spin-orbital order}} \\
    $x$                & \multicolumn{2}{c}{0}    &   \multicolumn{2}{c}{0.0625} \\
\hline
 $E_{\rm HF}$ (eV)     & \multicolumn{2}{c}{123.507} &  \multicolumn{2}{c}{122.156} \\
 $G$ (eV)              & \multicolumn{2}{c}{4.76}    &  \multicolumn{2}{c}{0.23}    \\
\hline
                                &    \multicolumn{4}{c}{Mn ion sublattice}    \\
electron density                &  $m=1$   &  $m=2$  &  $m=1$   &  $m=2$      \cr
$\langle n_{m,xy} \rangle   $   &   1.02   &   1.02  &   1.02   &   1.02      \cr
$\langle n_{m,yz} \rangle   $   &   1.00   &   1.02  &   1.00   &   1.02      \cr
$\langle n_{m,zx} \rangle   $   &   1.02   &   1.00  &   1.02   &   1.00      \cr \hline
$\langle n_{m,x^2-z^2} \rangle$ & \textbf{0.12} & 0.88 & \textbf{0.11} & 0.80 \cr
$\langle n_{m,3y^2-r^2}\rangle$ & \textbf{1.00} & 0.24 & \textbf{0.95} & 0.27 \cr
\hline
$\langle n_{m,y^2-z^2} \rangle$ & 0.89 & \textbf{0.12} & 0.81 & \textbf{0.11} \cr
$\langle n_{m,3x^2-r^2}\rangle$ & 0.23 & \textbf{1.00} & 0.26 & \textbf{0.96} \cr
\hline
$\langle n_{m,x^2-y^2} \rangle$ &   0.67   &   0.68   &   0.68   &   0.69     \cr
$\langle n_{m,3z^2-r^2}\rangle$ &   0.45   &   0.44   &   0.39   &   0.37     \cr
\end{tabular}
\end{ruledtabular}
\end{table}

With these results for the electron distribution, one could say that the
experimental facts are faithfully reproduced. However, it is not so, as
the HOMO-LUMO gap $G$ we obtained is 4.76 eV, much larger than the
experimentally measured; the experimental data concerning band-gap are
in the range of $1.1-1.7$ eV (direct gap) \cite{ari93,sai95,tob01} and
0.24 eV \cite{mah95} (indirect  gap).
In fact, the HOMO-LUMO gap should correspond either to direct or to
indirect gap, whichever is smaller. Anyway, for these both
possibilities this discrepancy is by far too large and it invites one
to reject thinking in terms of ideal-ionic model and to consider
instead non-zero self-dopings $x$. Note that our cluster is rather
small thus we can not (in the following) consider and study any
arbitrary value of self-doping $x$ as this would in some cases
result in non-integer electron number (in the cluster), and is some
other case would result in an open-shell system (and such systems
can not represent the ground state of an infinite crystal).

The density distribution found without self-doping (at $x=0$) suggests
that there is some but rather small electron transfer from O($2p$) to
Mn($3d$) orbitals, see Table II. Indeed, we have verified that the
electron density in all oxygen $p$-orbitals is very close to 6.0, thus
we deal with an almost ``perfect'' ionic system with O$^{-2}$ ions.

\subsection{Weak self-doping $x=1/16$}
\label{sec:x=1/16}

The first possible close shell configuration is obtained by
subtracting only 4 electrons from the total electron number in the
$4\times 4\times 4$ cluster with $N_{\rm el}=1408$ electrons at $x=0$.
In this case we obtain the orbital order and spin order virtually the
same as for $x=0$ case (compare the density distribution for $x=0$ and
$x=0.0625$ in Table II). However, total electron density in $e_g$
orbitals is reduced by self-doping, see Table II.
But most importantly, the HOMO-LUMO gap $G$ becomes much reduced to
0.23 eV, in satisfactory (though probably incidental) agreement with
the experimental results \cite{mah95}.

Once again, one finds the ground state with the $x$ quantization axis
for the magnetization, and the complementary $A$-AF phase with the $z$
easy spin direction is by 1.5 meV higher (per one unit cell). Other
magnetic states are less favorable. The state with FM order is only by
2.0 meV higher, and the states with $G$-AF and $C$-AF spin order are
both by $\sim 26$ meV higher than the ground state. This result is
important and reflects the proximity of the FM order in doped systems,
which can be stabilized at still higher doping $x\sim 0.17$, as known
from the phase diagram \cite{Cheon}. In fact, in the $A$-AF ground state
the exchange interactions in $ab$ planes are FM, and the described
change of spin order involves just the change of sign of the
exchange interaction along the $c$ axis, from AF to FM.

\subsection{Moderate self-doping: $x=1/8$}
\label{sec:x=1/8}

Computations for $x\ge 1/8$ invariably produce the states with zero
(or negative) HOMO-LUMO gaps. This can serve as an indication that the
FM metallic regime sets in already at this self-doping level in the
cluster under consideration.

Note that the experimental results for La$_{1-x}$Sr$_x$MnO$_3$ systems
indicate that such systems are conducting and FM for $x>0.2$ doping
\cite{Cheon}. If we roughly identify our theoretical value of
self-doping $x=1/8\approx 0.12$ with the doping by Sr, we approach the
metallic regime, even when this value of $x$ does not coincide yet with
the experimental doping in metallic FM manganites ($x\sim 0.17$)
\cite{end99,iva02}. We remark that such a discrepancy for small
(not infinite) cluster and for simple non-\textit{ab-initio} $d-p$ model
can be expected.

\subsection{Neglecting splitting within $e_g$ and $t_{2g}$ states}
\label{sec:nos}

As already discussed, the modeling of JT Hamiltonian goes in two
separate steps:
(i) changing the bare energies of individual orbitals, and
(ii) performing Harrison \cite{Har05} scaling of hopping integrals due
to modified Mn-O bond lengths \cite{sna18}, i.e., changing
simultaneously static crystal-field potential and changing the kinetic
part.
To the best of our knowledge, only the second step (ii) is discussed
in the literature.

Therefore (to conform to the main stream) we performed auxiliary
computations putting to zero bare level splittings within $e_g$ and
$t_{2g}$ multiplets but performing Harrison scaling to adjust the
values of $d-p$ hopping elements to the actual bond lengths. It could
appear surprising, but the results concerning the ground states did not
change much. The magnetic order and orbital order persist, albeit the
orbital order is somewhat weaker.
Thus it seems that the change which the JT-effect brings upon kinetic
Hamiltonian part (hopping elements) is the dominant change or at least
it is just enough for a satisfactory modeling of the JT effect.

\section{ROBUSTNESS OF ORBITAL ORDER}
\label{sec:oo}

Finally, we address the question of stability of orbital order under
self-doping. In doped manganites orbital order persists at low doping
up to $x\simeq 0.1$ \cite{Tok06}, and at higher doping orbital liquid
\cite{Fei05} takes over which supports FM metallic phase. A remarkable
feature of the perovskite vanadates is that orbital order is quite
robust \cite{Miy00,Fuj05} and is destroyed only at high concentration
of charged defects $x\simeq 0.20$ by orbital polarization interactions
which frustrates orbital order \cite{Ave19}.

\begin{table}[b!]
\caption{The HOMO-LUMO gap $G$, the average magnetization value,
$|\langle S^x\rangle|$, and the orbital order parameters, $\eta_{y/x}$
and $\eta_{z}$, versus hypothetical values of self-doping $x$ for
LaMnO$_3$. All the presented data correspond to ground states
characterized by coexisting $A$-AF spin (with $x$-easy axis of
magnetization) and $C$-AO order.
}
\begin{ruledtabular}
\begin{tabular}{ccccc}
 $x$   & $G$ (eV) & $|\langle S^x\rangle|$ & $\eta_{y/x}$ & $\eta_z$ \\ \hline
$0$    &   4.76     &       1.98      &        0.79       & $-0.20$  \\

$1/16$ &   0.23     &       1.95      &        0.79       & $-0.28$  \\
$2/16$ & $\simeq 0$ &       1.92      &        0.78       & $-0.32$  \\
$3/16$ & $\simeq 0$ &       1.89      &        0.77       & $-0.35$  \\
\end{tabular}
\end{ruledtabular}
\label{tab:oo}
\end{table}

To investigate orbital order and its dependence on self-doping $x$,
we take $N_e=64\times(22-x)$ electrons for
\mbox{$(4\times 4\times 4)$-cluster}. Having in mind the charge
distribution anisotropy in distorted rhombuses in $ab$ planes, we
define orbital order parameter $\eta_{y/x}$ for the observed $C$-AO
order as follows,
\begin{eqnarray}
\label{yx}
2 \eta_{y/x}&=&
   \frac{ -\langle n_{1,x^2-z^2}\rangle + \langle n_{1,3y^2-r^2}\rangle}
        {  \langle n_{1,x^2-z^2}\rangle + \langle n_{1,3y^2-r^2}\rangle} \nonumber \\
&+&\frac{ -\langle n_{2,y^2-z^2}\rangle + \langle n_{2,3x^2-r^2}\rangle}
        {  \langle n_{2,y^2-z^2}\rangle + \langle n_{2,3x^2-r^2}\rangle}\,,
\end{eqnarray}
where the electron densities appropriate to $y/x$ quantization directions
[as applied to $(m=1)$ / $(m=2)$ rhombuses; see Fig. 1] are used.

Once again we stress that when only one $z$-quantization direction (on
all rhombuses) is used, then the $C$-AO orbital order is not visible at
all. To make this picture complete we can define ordinary order
parameter $\eta_z$ as well,
\begin{eqnarray}
\label{etaz}
2 \eta_z &=&
   \frac{ -\langle n_{1,x^2-y^2}\rangle + \langle n_{1,3z^2-r^2}\rangle}
        {  \langle n_{1,x^2-y^2}\rangle + \langle n_{1,3z^2-r^2}\rangle} \nonumber \\
&+&\frac{ -\langle n_{2,x^2-y^2}\rangle + \langle n_{2,3z^2-r^2}\rangle}
        {  \langle n_{2,x^2-y^2}\rangle + \langle n_{2,3z^2-r^2}\rangle}\,,
\end{eqnarray}
which only shows the difference between
electron densities $\langle n_{m,x^2-y^2}\rangle$ and
$\langle n_{m,3z^2-r^2}\rangle$, (influenced by $z$-direction tetragonal
distortion) which is \textit{independent} of $m$, i.e., they are the
same for ($m=1$)-type and for ($m=2$)-type rhombuses. We remark that
sometimes one finds a tiny site-to-site charge modulation which is
possibly an artifact due to imperfect convergence, we neglect it --- to
get rid of it in Eq. (\ref{etaz}) we average $\eta_z$ over $m=1$ and
$m=2$ rhombuses.

The $C$-AO order parameter (\ref{yx}) versus hypothetical values of $x$
(self-dopings) is shown in Table \ref{tab:oo}. It is very robust and
almost independent of self-doping $x$ up to rather high value $x=3/16$.
It is remarkable that $C$-AO order survives in the metallic regime at
$x=2/16$ and $x=3/16$. On the contrary, the parameter $\eta_z$
(\ref{etaz}) is inconclusive concerning $C$-AO order. Instead, it shows
that the electron density in $x^2-y^2$ orbitals is higher than the one
in $3z^2-r^2$, in agreement with the model including tetragonal
distortions \cite{sna18}.

\section{Summary and conclusions}
\label{sec:summa}

We have shown that the $d-p$ model with strong electron interactions
reproduces correctly spin-orbital order in LaMnO$_3$, provided the
electronic configuration of Mn ions is very close to Mn$^{3+}$ and
the oxygen distortions due to the Jahn-Teller effect are included.
This implies also selecting the adequate orbital basis which is the
most appropriate to describe the orbital ordered state stabilized by
oxygen distortions. Thereby
occupied $e_g$ orbitals follow the oxygen distortions in $ab$ planes,
and one finds $A$-AF / $C$-AO order, as observed \cite{Dag01,Cheon}.

We have shown that the self-doping in LaMnO$_3$ is small but finite and
is in fact necessary to reproduce the observed insulating behavior with
a small gap. This result emphasizes the importance of electronic charge
delocalization over O($2p$) orbitals in the $d-p$ model for a
charge-transfer insulator LaMnO$_3$.

This study completes the series of papers
\cite{Ros15,Ros16,Ros17,Ros18}, where we have shown that the multiband
$d-p$ model is capable of reproducing coexisting spin-orbital order in
various situations and in various perovskites. In contrast to
time-consuming \textit{cluster-ab-initio} or LDA+$U$ calculations
\cite{yin06,yam06,kov11,gon11,mel15}, the computations using $d-p$ model
are very efficient and should be regarded as easy and simple tool for
any preliminary study to establish the electronic structure and ground
state properties of an investigated perovskite. During such calculations
the only difficult part is the proper choice of the Hamiltonian
parameters. We suggest that this approach could be a promising technique
to investigate heterostructures \cite{Yu09,Rog12} or superlattices
\cite{Cao11,Don12}, where the Jahn-Teller effect plays an important
role. On the other hand, when information about the correct values of
Hamiltonian parameters are uncertain, one can perform computations
with several sets of Hamiltonian parameters. The results of such
computations when confronted with the experimental results could be
eventually used for screening out wrongly chosen Hamiltonian
parameter sets.

\acknowledgments
We kindly acknowledge Narodowe Centrum Nauki (NCN, Poland)
Project No.~2016/23/B/ST3/00839.
A.~M.~O. is also grateful for the Alexander von Humboldt
Foundation Fellowship (Humboldt-Forschungspreis).

\end{document}